\documentclass{article}
\usepackage{graphicx}
\usepackage{dcolumn}
\usepackage{bm}

\def\be{\begin{equation}}
\def\ee{\end{equation}}
\def\ba{\begin{array}}
\def\ea{\end{array}}
\def\bea{\begin{eqnarray}}
\def\eea{\end{eqnarray}}

\begin{document}

\title{Momentum Dependence of Nuclear Mean Field and multifragmentation in Heavy-Ion Collisions}
\author{Yogesh K. Vermani, Supriya Goyal and Rajeev K. Puri
\footnote{rkpuri@pu.ac.in} \\
Department of Physics, Panjab University, \\
Chandigarh-160014, India.}

\maketitle

\begin{abstract}

We report the consequences of implementing momentum dependent
interactions (MDI) on multifragmentation in heavy-ion reactions
over entire collision geometry. The evolution of a single cold
nucleus using static soft equation of state and soft momentum
dependent equation of state demonstrates that inclusion of
momentum dependence increases the emission of free nucleons.
However, no heavier fragments are emitted artificially. The
calculations performed within the framework of \emph{quantum
molecular dynamics} approach suggest that MDI strongly influence
the system size dependence of fragment production. A comparison
with ALADiN experimental data justifies the use of momentum
dependent interactions in heavy-ion collisions.
\end{abstract}

\section{Introduction}

The heavy-ion collisions have always played a fascinating role in
exploring various aspects of nuclear dynamics such as
fission-fusion \cite {ono, bruc,wang}, multifragmentation
\cite{aich, aich91, peil, peil1, jai, reis, blaich, schut, poch},
collective flow \cite{peil, reis, moli, mages, sood, soff, sood1}
as well as particle production \cite{aich91, rose, hart} etc. The
wider energy spectrum available due to modern accelerator
technologies has become a powerful tool in experimental and
theoretical studies in exploring the nature of hot and compressed
nuclear medium via heavy-ion collisions.

It is well accepted that the outcome of a reaction depends not
only on the density but also on the momentum space \cite{rose,
khoa}. The momentum dependence of the equation of state can be
extracted from the real part of the optical potential. This
potential is expected to affect those nucleons from target and
projectile which possess larger relative momenta. The momentum
dependence of the nucleon-nucleon (\emph{n-n}) potential is found
to affect drastically the collective flow observables \cite{breng,
sk} and particle production \cite{aich, rose, hart}. It has been
shown that observables related to the particle production
\emph{e.g.} $\pi, \kappa, \lambda$ yields, $n_{d}/n_{p}$ ratios
etc. are strongly influenced by the momentum dependence of the
\emph{n-n} interaction \cite{rose, hart}. A strong influence of
momentum dependent interactions was also observed on fragment flow
for $E_{lab}\geq 400~ MeV/nucleon$. In higher incident energy
regime, momentum dependent interactions (MDI) cause stronger
reduction in the number of \emph{n-n} collisions leading to more
pronounced transverse flow. The momentum dependent interactions,
therefore, increase the mean free path of nucleons, and
consequently affect the stopping and thermalization of nuclear
matter \cite{breng, jia}.

On the contrary, very few attempts exist in the literature, that
shed light on the consequences of implementing momentum dependent
interactions in fragmentation \cite{blaich, sk}. One of the basic
problem with MDI is the strong repulsion created in the nuclear
environment. As a result, nuclei propagating with MDI tend to be
destabilized and start decaying via emission of free nucleons and
clusters quite early during the reaction. Interestingly, the role
of momentum dependent interactions depends crucially on the impact
parameter of the reaction. It is found to enhance the energy of
disappearance of flow in central collisions \cite{sood}, whereas
it reduces the energy of disappearance of flow in peripheral
collisions \cite{soff, sood1}. Similarly, MDI reduce the
production of fragments in central collisions whereas it enhances
the same in peripheral collisions \cite{sk}. However, one always
remained concerned about the stability of nuclei propagating with
momentum dependent interactions. Even a use of cooling procedure
via Pauli potential is also reported in the literature
\cite{peil1}. Our present aim, therefore, is to investigate the
stability of cold nuclei propagating under the influence of
momentum dependent interactions and to see whether one can study
fragmentation with MDI or not. An attempt to study the system size
effects in the presence of momentum dependent forces will also be
made. We shall also confront our calculations with
multifragmentation data of ALADiN group \cite{schut} which has a
\emph{rise and fall} variation with impact parameter. This study
is carried within the framework of quantum molecular dynamics
(QMD) transport model \cite{aich, aich91}. The QMD model and
implementation of momentum dependent potential are described in
section \ref{model}. Our results are presented in section
\ref{results} and summarized in section \ref{summary}

\section {\label{model} QMD Model and Momentum Dependent Interactions}

The quantum molecular dynamics (QMD) model \cite{aich,hart} is an
n-body transport theory that simulates the heavy-ion (HI)
reactions between 30 MeV/nucleon and 1 GeV/nucleon on event by
event basis. It includes quantum features like Pauli blocking,
stochastic scattering and particle production. Here each nucleon
is represented by Wigner distribution function of the form:
\begin{equation}
f_{i} ({\bf r,p,t})=\frac{1}{(\pi \hbar)^3} e^{-({\bf r}-{\bf
r}_i(t))^2/2L}. e^{-({\bf p}- {\bf p}_i(t))^2 2L/{\hbar}^2},
\end{equation}
where $\bf{r}_{i}(t)$ and $\bf{p}_{i}(t)$ define the classical
orbit, the center of $i^{th}$ Gaussian wave packet in phase space
which evolves in time. The centers of these Gaussian wave packets
propagate according to the classical equations of motion
\cite{aich, peil}. The interaction part used in the QMD model
consists of local Skyrme interaction, a finite range Yukawa term
and an effective Coulomb interaction among protons \emph{i.e.}
\begin{equation}
 V^{tot}= V^{Sk} + V^{Yuk}+ V^{Coul},\label{tot}
\end{equation}
with local Skyrme interaction consisting of two- and three-body
interactions:
\begin{equation}
V^{Sk}=t_{1}\delta({\bf r}_i-{\bf r}_j) + t_{2}\delta({\bf
r}_i-{\bf r}_j) \delta({\bf r}_i-{\bf r}_k). \label{sk}
\end{equation}
\noindent Since QMD model is a \emph{n}-body theory, therefore,
Eq. (\ref{sk}) can be reduced to a density dependent potential in
the limit of infinite nuclear matter limit as:
\begin{equation}
U^{Sk} =\alpha (\frac{\rho }{\rho _{o} })+\beta \left(\frac{\rho
}{\rho _{o} } \right)^{2} . \label{sky}
\end{equation}
\noindent The momentum dependence of nuclear mean field is
included in Eq. (\ref{tot}) via momentum dependent interaction as:
\begin{equation}
V^{MDI}=t_{4}ln^{2}\left[t_{5} \left({\bf p}_{i} -{\bf p}_{j}
\right)^{2} +1 \right]\, \delta \left({\bf r}_{i} -{\bf r}_{j}
\right), \label{md}
\end{equation}
\noindent with parameters $t_{4}= 1.57~MeV$ and $t_{5}=5.0 \times
10^{-4}~(MeV/c)^{-2}$. This parameterization is deduced from the
real part of the proton-nucleus optical potential which reproduces
the experimental data upto 1 GeV/nucleon \cite{aich, hama}. In an
infinite nuclear matter limit, generalized \emph{n-n} potential
(Eq.(\ref{tot})~and (\ref{md})) leads to following density and
momentum dependent potential (without Coulomb and Yukawa terms):
\begin{equation}
U(\rho ,{\bf p})=\alpha \left({\frac {\rho}{\rho_{0}}}\right) +
\beta \left({\frac {\rho}{\rho_{0}}}\right)^{\gamma}+t_{4}ln^{2}
[t_{5} ({\bf p}^{2} +1)]\left(\frac{\rho }{\rho _{o} } \right).
\label{mdi}
\end{equation}
The parameters $\alpha$, $\beta$ and $\gamma$ in Eq.(\ref{mdi})
have to be re-adjusted in the presence of momentum dependent
interactions so as to reproduce the ground state properties of
nuclear matter. The parameters corresponding to soft, hard and
their momentum dependent versions are labelled as S, H, SM and HM,
respectively. The constants $\alpha$ and $\beta$ give the proper
rms radii and binding energies of nuclei across the periodic
table. Parameter $\gamma$ gives us possibility to examine the
compressibility and hence equation of state. The set of parameters
corresponding to soft (S), hard (H) and their momentum dependent
versions SM and HM, respectively, can be found in Ref.
\cite{aich91}.

\section{\label{results} Results and discussion}

\subsection{Stability of cold QMD nuclei}
To address the question of stability of computational nucleus in
the presence of momentum dependent interactions (MDI), we
initialize a single cold projectile using soft (S) equation of
state and soft momentum dependent equation of state (SM). Earlier
theoretical attempts ranging from giant monopole resonances
\cite{itoh} to nucleosynthesis of heavy elements in mergers of
neutron stars \cite{ross} could be explained if the equation of
state (EoS) is relatively \emph{soft} than when it is
\emph{stiff}. Another study concerning the linear momentum
transfer occuring in central HI collisions also showed that a soft
compressibility modulus is needed to explain the experimental data
\cite{cib, hdad}. These observations motivated us for the choice
of comparatively softer EoS. We follow the cluster emission
pattern and rms radii of few computational nuclei. Figure 1 shows
the time evolution of cold QMD nuclei, namely $^{58}Ni$ and
$^{197}Au$ initialized with S and SM interactions. The cluster
emission is followed for the time span of 200 fm/c. Here,
$A^{max}$ denotes the size of residual nucleus. This should be
close to that of parent nucleus if there is no destabilization of
the nucleus. \nolinebreak
\begin{figure}[!t]
\centering
\includegraphics[scale=0.5, trim=16 0 0 0] {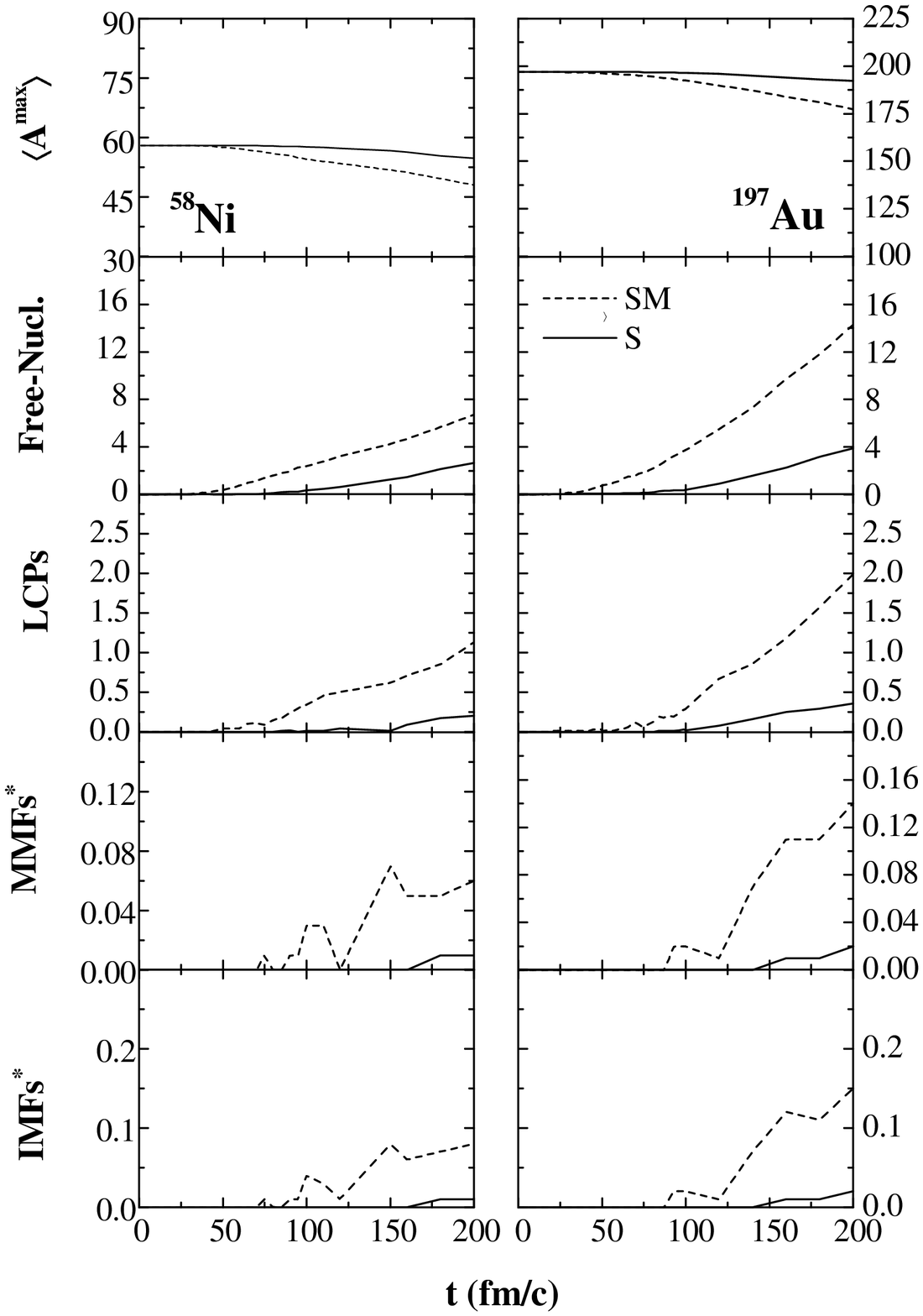}
\vskip -0.8cm \caption {The time evolution of heaviest fragment
$\langle A^{max} \rangle $, free nucleons, LCPs $[2\leq A \leq
4]$, $MMFs^{*}$ $[5\leq A \leq 9]$ and $IMFs^{*}$ $[5\leq A \leq
A_{P}/3]$ ($A_{P}$ being the mass of projectile nucleus) emitted
from a single cold nucleus of $^{58}Ni$ (left panel) and
$^{197}Au$ (right panel). Results obtained with soft (S) equation
of state are represented by solid lines whereas dashed lines show
results with soft momentum dependent (SM) equation of state. `*'
indicates that heaviest fragment has been excluded.}
\end{figure}
The sizes of parent $^{58}Ni$ and $^{197}Au$ nuclei reduce with
the inclusion of momentum dependent interactions compared to
nuclei propagating with static soft interactions alone. The SM
interactions caused an enhanced emission of free nucleons and
light charged particles LCPs $[2\leq A \leq 4]$. However, medium
mass fragments MMFs $[5\leq A \leq 9]$ and IMFs \{$[5\leq A \leq
A_{P}/3];~A_{P}$ being the mass of projectile\} are almost
insensitive towards momentum dependent interactions. Superscript
(*) indicates that heaviest fragment has been excluded from the
multiplicities of MMFs and IMFs. Only a small fraction is emitted
as intermediate mass fragments (IMFs). It seems that nucleons
close in space are emitted in bulk, therefore, leading to an
enhanced emission of light clusters. On the contrary, very few
nucleons, LCPs and heavier clusters are emitted when propagating
with soft EoS. The enhanced evaporation with MDI is also due to
repulsive nature of these interactions. Does this enhanced
emission prohibit one to use MDI for fragmentation ? If one sees
carefully, majority of mass that leaves the gold nucleus (for
example, with MDI about 19 units are emitted and $\langle A^{max}
\rangle$ is close to 177) is in the form of free nucleons. In the
above gold nucleus, out of 19 units about 15 are in terms of free
nucleons. In other words, we see that nucleons from the surface
are emitted and there is no contribution towards the emission of
intermediate mass fragments. One sees that even with MDI, only
0.15 IMFs are emitted on the average. Realizing that as many as
10-12 IMFs can be seen emitted in Au+Au reaction \cite{tsang},
this number with MDI is negligible.
\begin{figure}[!t]
\centering \vskip -1cm
\includegraphics[scale=0.45] {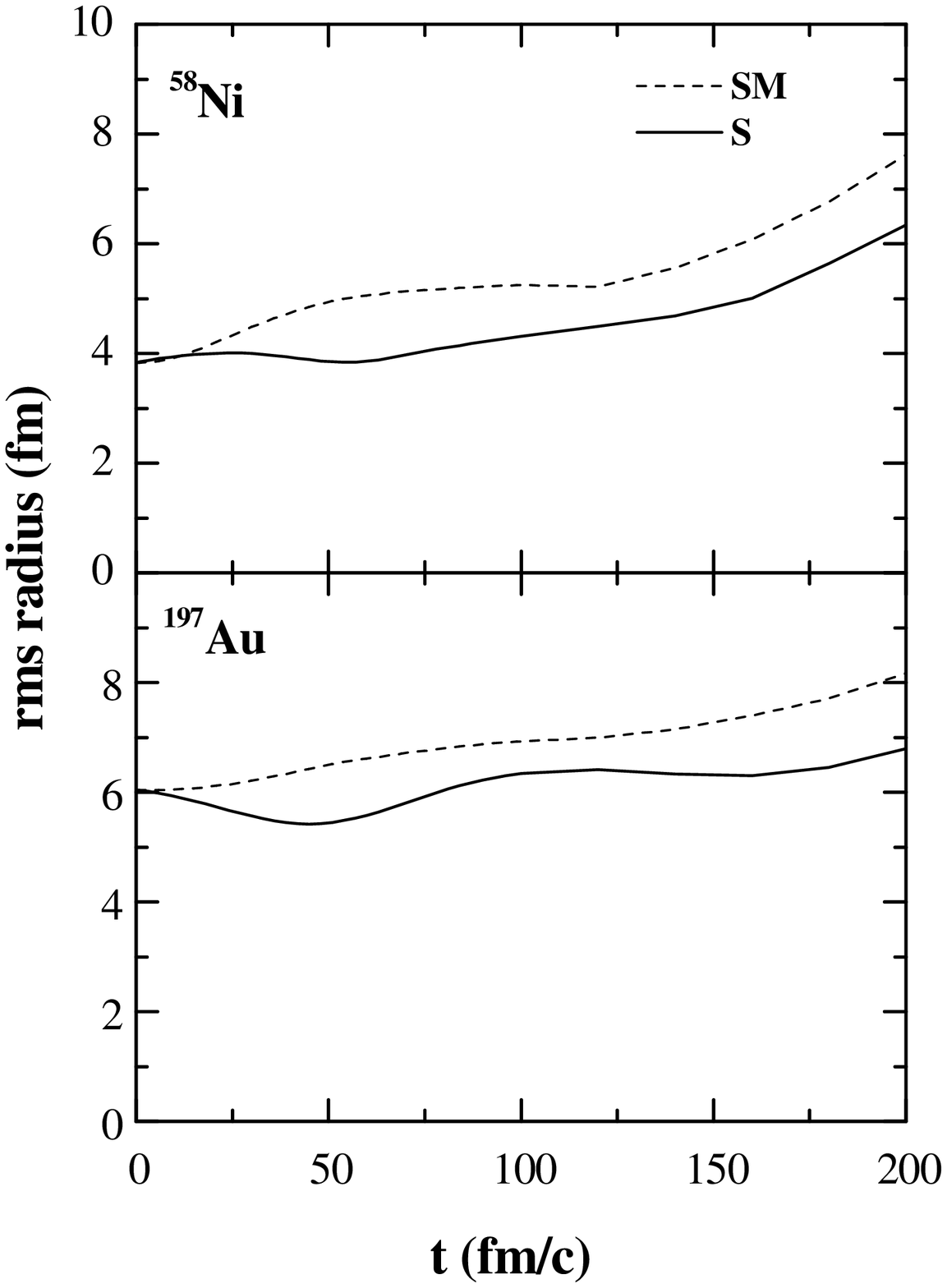}
\vskip -1.25cm \caption {The time variation of rms radii of single
cold nuclei of $^{58}Ni$ (top panel) and $^{197}Au$ (bottom panel)
using soft (S) equation of state (solid lines) and soft momentum
dependent (SM) interactions (dashed lines).}
\end{figure}

\begin{figure}[!t]
\centering \vskip -0.66cm
\includegraphics*[scale=0.45] {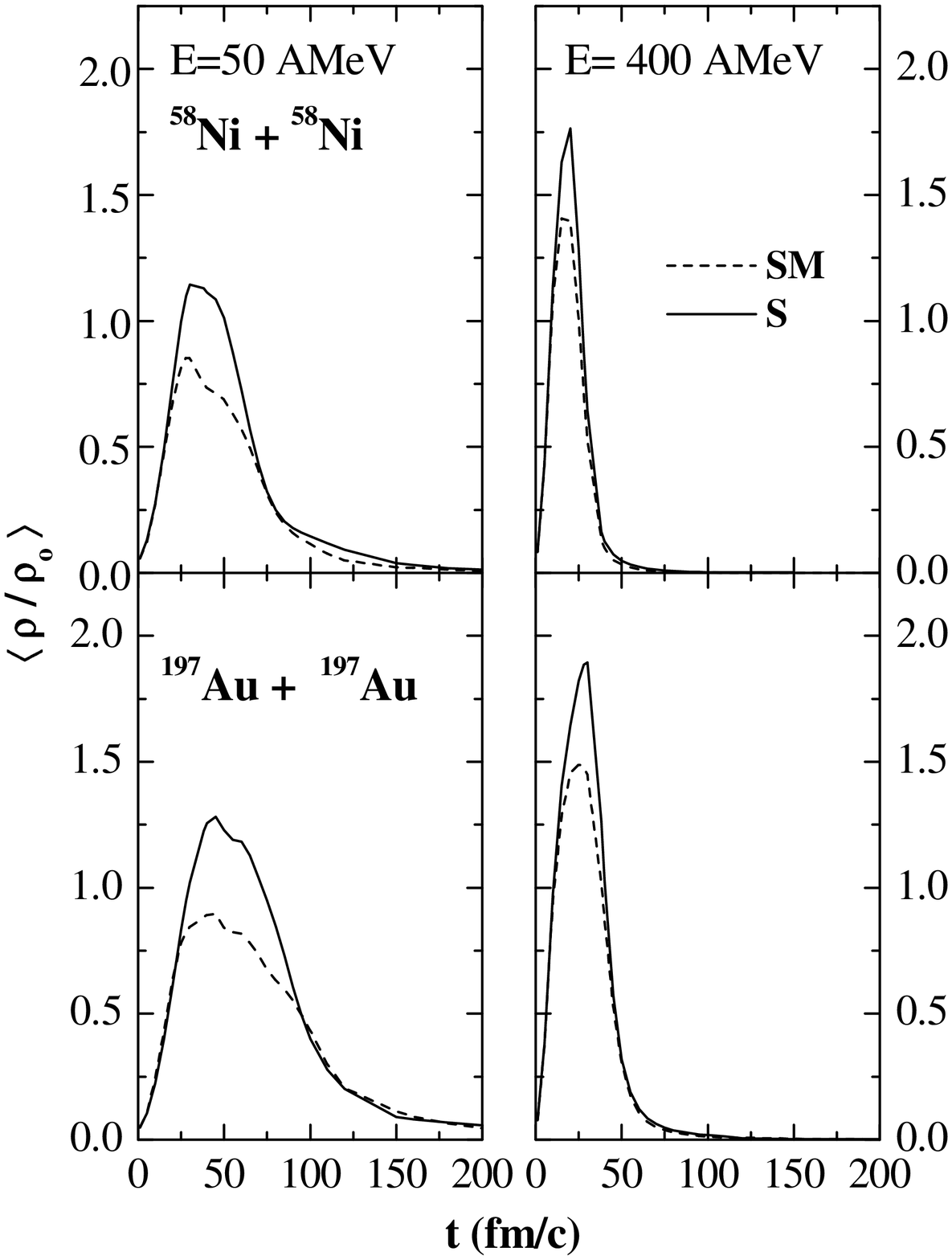}
\vskip -0.80cm \caption {The average nucleonic density $\langle
\rho/\rho_{o} \rangle$ calculated in a central sphere of 2 fm
radius versus reaction time for the central collisions of
$^{58}Ni+^{58}Ni$ (top panel) and $^{197}Au+^{197}Au$ (bottom
panel). The results obtained with soft (S) and soft momentum
dependent (SM) interactions are compared at 50 AMeV (left) and 400
AMeV (right).}
\end{figure}
A survey of the time evolution of  rms radii of a single QMD
nucleus also depicts the same picture. We show in Fig. 2, the time
evolution of rms radii of $^{58}Ni$ and $^{197}Au$ nuclei till 200
fm/c. The rms radius of nucleus with SM interactions increases
gradually compared to that initialized with static soft
interactions. This behavior reflects that MDI create additional
repulsions among nucleons which leads to enhanced emission of free
nucleons. The rms radii of gold and nickel nuclei in soft case
shows negligible deviation for the characteristic time of HI
collision. As discussed above, this enhanced radius is due to the
emission of free nucleons and not due to the IMFs. Therefore, one
can study the fragmentation with MDI since the structure of IMFs
is not altered by the inclusion of MDI.

\subsection{Heavy-ion collisions and system size effects}

After investigating the behavior of cold nuclei initialized with
momentum dependent interactions, let us study the effect of
momentum dependent forces in heavy-ion reactions. One of the
observables linked with the compression and expansion of nuclear
matter is the density of fragmenting system. The total nuclear
matter density is obtained as :
\begin{equation}
\rho (\bf{r},t) = \sum_{j=1}^{A_{T}+A_{P}}\frac{1}{(2\pi L)^{3/2}}
e^{-(\bf{r}-\bf{r}_{j}(t))^{2}/2L}.
\end{equation}
Here $A_{T}$ and $A_{P}$ stand for the target and projectile
masses, respectively. In our approach, average nuclear matter
density $\langle\rho/\rho_{o}\rangle$ is calculated in a sphere of
2 fm radius.

In Fig. 3, we display the time evolution of average nucleon
density $\langle\rho/\rho_{o}\rangle$ reached in the central
region for the head-on collisions  of $^{58}Ni+ ^{58}Ni$ and
$^{197}Au+ ^{197}Au$ at incident energies of 50 and 400 AMeV. The
maximal average density tends to reduce with inclusion of momentum
dependent interactions. This happens due to additional \emph{n-n}
repulsions created in the system that prohibits compression of
nuclear matter to a significant level. This difference in the
behavior of $\langle\rho/\rho_{o}\rangle$ calculated using S and
SM interactions diminishes at higher incident energies (400 AMeV).
This is due to the fact that in central collisions at 400 AMeV,
most of the initial \emph{n-n} correlations are already destroyed
and matter is already scattered, and therefore, repulsion
generated due to MDI does not play any significant role. As a
result, we do not see much difference in average central density
reached at higher incident energy. Another important quantity
related with the initial compression of nuclear matter is the rate
of binary collisions. We have also checked the collision rate for
these reactions and its behavior is found to be consistent with
the density profile.
\begin{figure}
\centering \vskip -0.7cm
\includegraphics*[scale=0.45] {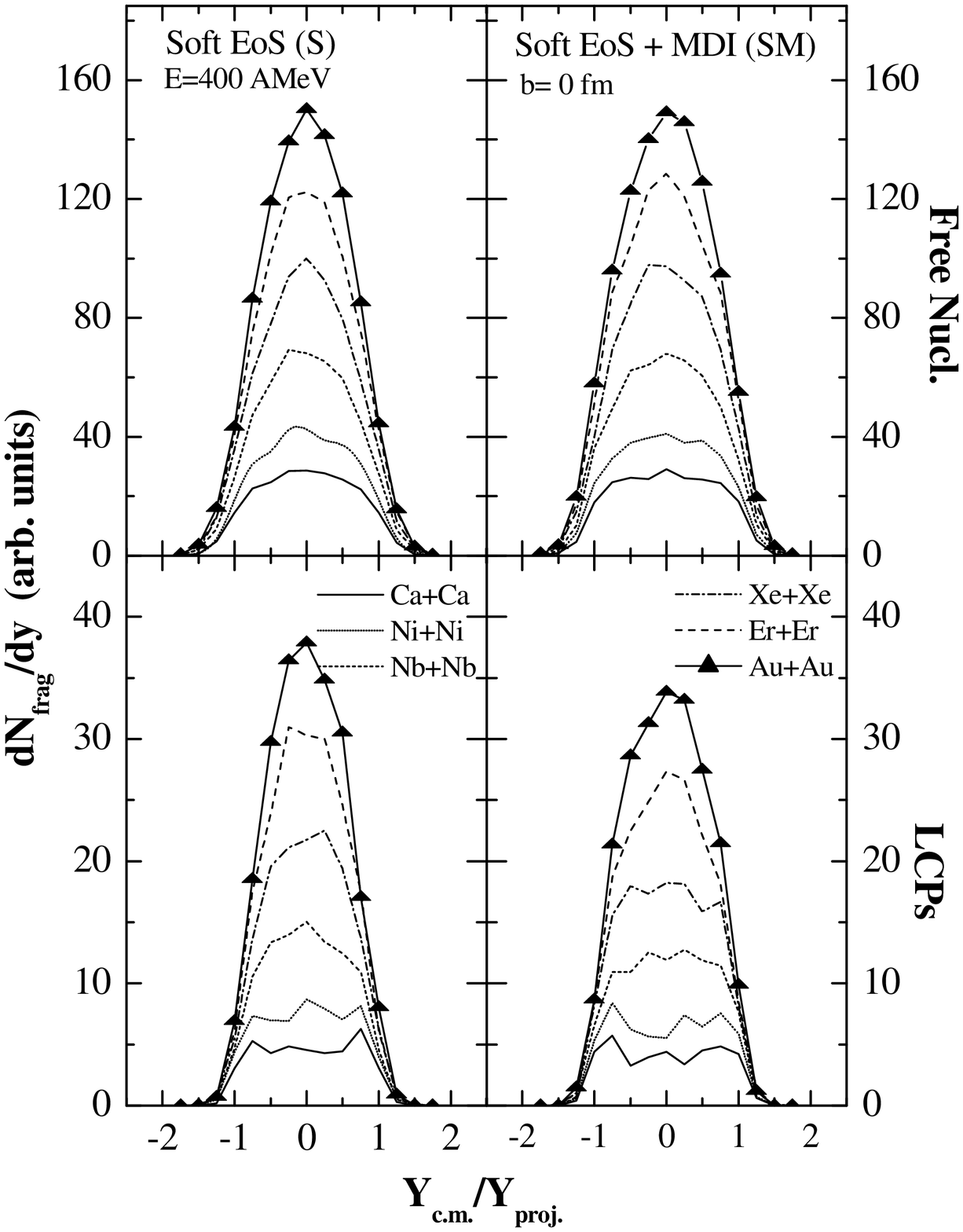}
\vskip -0.8cm \caption {The rapidity distribution $dN_{frag}/dy$
of free nucleons and LCPs $[2\leq A \leq 4]$ as a function of
scaled rapidity $Y_{c.m.}/Y_{proj.}$; $Y_{proj.}$ being the
rapidity of projectile for the head-on collisions at 400 AMeV.}
\end{figure}
The rapidity distribution of nucleons is another useful tool to
characterize the stopping and thermalization of the nuclear
matter. We have displayed in Fig. 4, the fragment rapidity
distribution $dN_{frag}/dy$ of free nucleons and LCPs for central
collisions of six different symmetric systems at 400 AMeV. The
results are displayed here using soft EoS (left panel) and soft
EoS including MDI (right panel). The rapidity distribution is more
`isotropic' and nearly full stopping is achieved in heavier
systems like Au+Au and Er+Er. In lighter systems, on other hand, a
larger fraction of particles is concentrated near target and
projectile rapidities resulting into broad Gaussian shape. This
feature can be seen in both S and SM cases. The lighter systems,
therefore, exhibit larger \emph{transparency} effect \emph{i.e.}
less stopping. Such features are also observed in the experimental
data of FOPI-group \cite{gwang}. Based on the experimental
observations and theoretical trends, one can say that smaller the
system, lesser is the stopping. With MDI, a slight increase in
transparency effect is seen due to lesser stopping of particles in
longitudinal direction. This happens due to reduction in
\emph{n-n} collisions which deflect the fragments in transverse
direction. As a results, one obtains less particles being stopped
in longitudinal direction.
\begin{figure}
\centering \vskip 0.6cm
\includegraphics[scale=0.45] {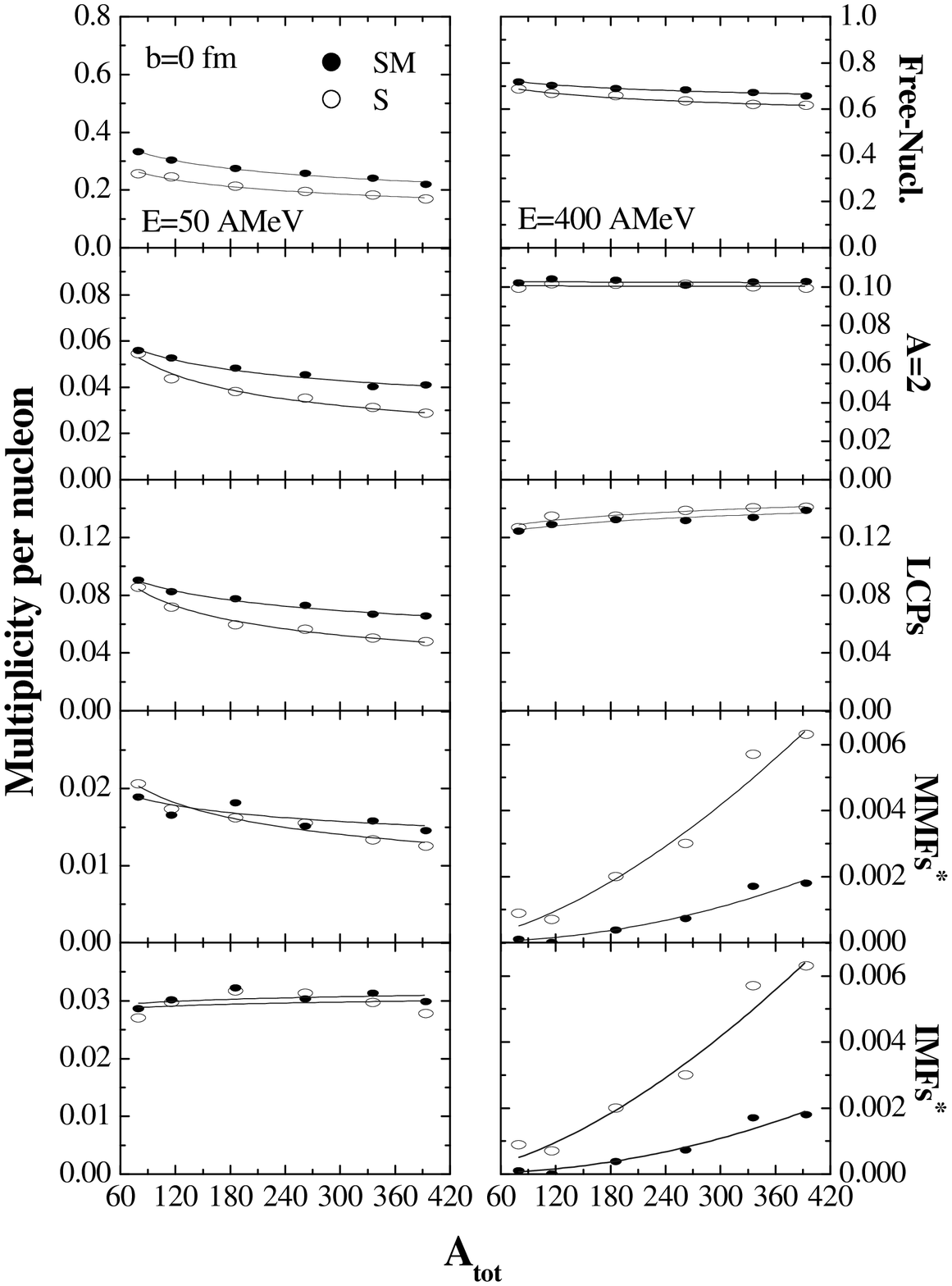}
\vskip -0.50cm \caption {The final state scaled multiplicity
(calculated at 200 fm/c) of free nucleons, fragments with mass
A=2, LCPs $[2\leq A \leq 4]$, $MMFs^{*}$ $[5\leq A \leq 9]$ and
$IMFs^{*}$ $[5\leq A \leq min \left\{ A_{P}/3,65 \right\}]$ as a
function of total mass of the system $A_{tot}$. Results shown here
are at incident energies of 50 AMeV (l.h.s) and 400 AMeV (r.h.s).
Open circles depict the calculations with soft (S) interaction
while solid circles are for soft momentum dependent (SM)
interactions. `*' means that heaviest fragment has been excluded.}
\end{figure}
The system size effects in the production probability of different
kinds of fragments has been studied and predicted by our group
\cite{jai}. Here we extend the same study with reference to
momentum dependent interactions. For this analysis, we simulated
the central collisions of six symmetric systems $^{40}Ca+
^{40}Ca$, $^{58}Ni+ ^{58}Ni$, $^{93}Nb+ ^{93}Nb$, $^{131}Xe+
^{131}Xe$, $^{168}Er+ ^{168}Er$ and $^{197}Au+ ^{197}Au$ at
incident energies of 50 and 400 AMeV. We also parameterized the
multiplicities as a function of total mass of the composite system
using a power law of the form: $cA_{tot}^{\tau}$; $A_{tot}$ being
the total mass of the system. Figure 5 displays the `reduced
multiplicity' i.e. multiplicity per nucleon of various kinds of
fragments. It is clear that the system size effects are more
visible in soft equation of state compared to soft momentum
dependent case. A negative slope obtained for the multiplicity of
free nucleons, fragments of mass A=2, and LCPs at 50 AMeV
indicates their origin from the surface of interacting nuclei. As
we move to momentum dependent version, additional break up of
\emph{n-n} correlations leads to enhanced emission of free
nucleons and light charged particles. As a result, multiplicity of
$MMFs^{*}$ $[5\leq A \leq 9]$ and $IMFs^{*}$ $[5\leq A \leq min
\left\{ A_{P}/3,65 \right\}]$ (excluding largest fragment
$A^{max}$) gets reduced at 400 AMeV, indicating the vanishing of
system size effect with MDI. In higher energy regime, cluster
production via emission of $MMFs^{*}$ and $IMFs^{*}$ is strongly
suppressed in the presence of MDI. It is worth mentioning that
earlier studies \emph{e.g.} see Ref. \cite{lie}, also reported the
momentum dependent potential to be more repulsive for high
momentum nucleons. This leads to enhanced emission of free
nucleons and LCPs. A similar enhancement of the nucleons emission
and light cluster production was predicted on inclusion of
momentum dependent effective interactions in the isoscalar nuclear
potential and symmetry potential \cite{lie, grec}. Contrary to
this, with static soft equation of state, the production
probability of MMFs and IMFs scale with the system size as power
law: $cA_{tot}^{\tau}$ with power factor $\tau$ close to 3/2.
\begin{figure}
\centering \vskip -0.7cm
\includegraphics[scale=0.45] {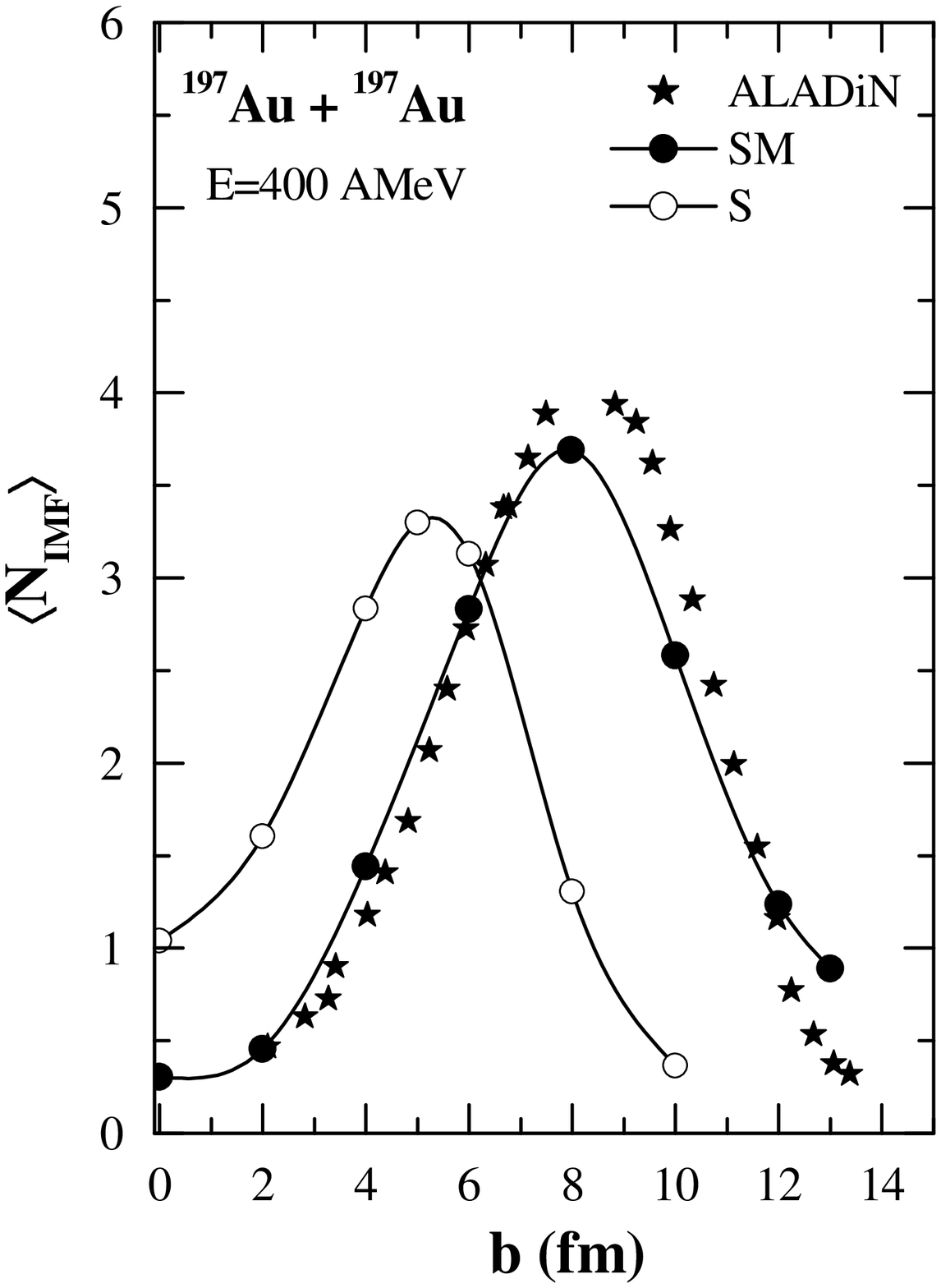}
\vskip -1.5cm \caption {The mean IMF multiplicity $\langle N_{IMF}
\rangle$ vs the impact parameter b for the reaction of $^{197}Au+
^{197}Au$ at 400 AMeV. The QMD calculations (at 300 fm/c) using
soft EoS (open circles) and soft momentum dependent EoS (solid
circles) are compared with ALADiN experimental data (filled
stars).}
\end{figure}
Let us now try to confront our calculations with experimental data
of ALADiN group \cite{schut}. The experimental data is very
fascinating because it has been shown that there is a \emph{rise
and fall} of multiplicity of intermediate mass fragments with
impact parameter \cite{schut}. However universality is observed
with mass of the system and with incident energies exceeding 400
AMeV. In Fig. 6, we display the multiplicity of intermediate mass
fragments as a function of impact parameter using soft (S) and
soft momentum dependent (SM) equations of state. We see that
entire  spectrum is very well reproduced by the momentum dependent
interactions. One should also keep in the mind that for central
impact parameters, different experimental groups like FOPI
\cite{reis}, ALADiN \cite{blaich, schut} and Miniball \cite{tsang}
differ significantly in the multiplicities of IMFs. Overall, we
see a clear need of momentum dependent interactions in heavy-ion
collisions.

\section{\label{summary}Summary}

Summarizing these findings, we here presented a detailed study on
the consequences of employing momentum dependent potential in
multifragment-emission. Investigation of a single cold nucleus
initialized with soft (S) and soft momentum dependent (SM)
equations of state reveals that momentum dependent interactions
act as \emph{destabilizing} factor, which results into enhanced
emission of free nucleons only. However, no change is seen towards
artificial emission of IMFs. Further, momentum dependent
interactions are observed to weaken the system size effects
studied at 50 and 400 AMeV.  A comparison of model calculations
with ALADiN data on Au+Au reactions favor strongly the use of
momentum dependent equation of state in heavy-ion collisions. \\

This work was supported by a research grant from Department of
Science and Technology, Government of India vide grant no.
SR/S2/HEP-28/2006.

\end{document}